\documentclass{aastex}
\usepackage{emulateapj5}
\usepackage{apjfonts}

\newcommand{\ea}{{\it et al.}}

\newcommand{\msol}{\mathrm{M}_\odot}

\newcommand{\kms}{km~$\rm{s}^{-1}$}

\newcommand{\beq}{\begin{equation}}
\newcommand{\eeq}{\end{equation}}
\newcommand{\bdm}{\begin{displaymath}}
\newcommand{\edm}{\end{displaymath}}

\shorttitle{Formation of Turbulent Cones in Accretion Disk Outflows}
\shortauthors{Poludnenko, Blackman, \& Frank}

\slugcomment{Submitted to the Astrophysical Journal Letters}

\begin{document}

\title{Formation of Turbulent Cones in Accretion Disk Outflows and 
Application to Broad Line Regions of Active Galactic Nuclei}

\author{A.Y. Poludnenko\altaffilmark{1}, E.G. Blackman\altaffilmark{2},
A. Frank\altaffilmark{3}}
\affil{Department of Physics and Astronomy,\\
       University of Rochester, Rochester, NY 14627-0171}
\altaffiltext{1}{wma@pas.rochester.edu}
\altaffiltext{2}{blackman@pas.rochester.edu}
\altaffiltext{3}{afrank@pas.rochester.edu}

\begin{abstract}
We consider the stability of an accretion disk wind to cloud formation
when subject to a central radiation force.  For a vertical launch
velocity profile that is Keplerian or flatter and the presence of a
significant radiation pressure, the wind flow streamlines cross in a
conical layer. We argue that such regions are highly unstable, and are
natural sites for supersonic turbulence and, consequently, density
compressions. We suggest that combined with thermal instability these
will all conspire to produce clouds. Such clouds can exist in
dynamical equilibrium, constantly dissipating and reforming.  As long
as there is an inner truncation radius to the wind, our model emerges
with a biconical structure similar to that inferred by Elvis (2000)
for the broad line region (BLR) of active galactic nuclei (AGN). Our
results may also apply to other disk-wind systems.
\end{abstract}


\section{INTRODUCTION}

Non-spherical outflows are a ubiquitous feature of AGN. In this Letter
we discuss the stability of such an outflow launched normally from the
disk with a decaying power law velocity profile. The motivation for
this work is to understand the conditions under which the flow may be
unstable to cloud formation, and to infer the structure of the region
in which such clouds may reside. \citet{Elvis} has argued that data
require AGN BLR clouds to reside in a narrow biconical structure. Our
work herein supports this possibilty.

A variety of launching mechanisms for AGN accretion disk outflows have
been studied: radiatively accelerated outflows \citep{Arav94},
hydrodynamic line-driven winds \citep{Proga99, Proga1},
hydromagnetic disk winds \citep{Konigl94}, \citep{Pudritz92}, thermal
wind-type outflows in low luminosity AGNs \citep{Pietrini}, and
others. In this work we parameterize the disk wind independent of the
launching mechanism.

A disk wind outflow can become unstable from linear and nonlinear
perturbations in velocity, density, temperature, ionization state,
etc. These can lead to cloud formation. The cloud model is a leading,
though not unanimously accepted, paradigm for the structure of the AGN
BLR regions: models exist with \citep{Elvis, Urry95} and without
\citep{Murray1, Murray2} clouds. The main reason for the lack of
agreement is the uncertainty about cloud survival in AGN enviroments
\citep{Mathews}. They may be unstable to evaporation (e.g., see 
\citep{Pier95}), or be insufficiently supported by pressure and be
dynamically unstable to shredding \citep{Mathews, Poludnenko}. Among
the suggested solutions has been magnetic confinement of clouds
\citep{Rees87, Bottorff1}.

However a key point is that a typical cloud needs only to exist long
enough to reprocess radiation \citep{Rees87, Celotti, Kuncic1,
Kuncic2}. If clouds are destroyed thereafter and new clouds are
formed, then the system can exist in dynamical equilibrium even though
each cloud loses its identity swiftly.  This may apply to the broad
line region. Short cloud survival and regeneration times can be
associated with turbulence \citep{Bottorff2, Bottorff3}. The balance
between formation and destruction maintains a constant BLR cloud
number density.

Our analysis suggests that a nonlinear wind instability resulting from
the combination of launching profile + radiation pressure can lead to
the formation of a turbulent biconical zone that might manifest itself
observationally as the BLR. This biconical structure is the region
where flow streamlines cross. Such a bicone is remarkably similar to
that inferred empirically by \citet{Elvis} and from numerical
simulations by \citet{Proga1}. Growing observational data
corroborates the existance of such a structure (e.g. NGC 1068
\citep{Arribas96}, Mrk 3 \citep{Ruiz01}, etc.)

We formulate the problem in section 2.1 and describe the flow field in
section 2.2. We analyze the linear and nonlinear stability of the flow
and a possible scenario for thermal instability resulting in formation
of the BLR clouds in section 2.3. In section 3 we discuss the results
and their implications for the AGN structure and dynamics.

\section{DISK WIND MODEL}

\subsection{Formulation of the problem}

All of our calculations are performed in cylindrically symmetric
geometry, however we do not include axial rotation. At the origin we
assume a black hole of mass $M_{BH}$. Fig.~\ref{streamlines} shows the
setup.

We take a Keplerian, geometrically thin, optically thick accretion
disk to be located along the abscissae with the distance $\lambda$
measured along the disk. The central engine is taken to be a
point-like source at the origin coincident with the black hole. We
exclude disk surface illumination by the central source including all
disk properties into the disk-wind launch velocity profile.

Our wind originates at the disk surface with launch velocity profile
\beq
v_{z0}(\lambda) = v^*\bigg(\frac{\lambda}{\lambda^*}\bigg)^n,
\label{vz0}
\eeq
where $v^*$ is the wind velocity at the innermost launch point of the
disk with coordinate $\lambda^*$ and $n$ is an arbitrary non-positive
number\footnote{We discard positive values of $n$ as unphysical.}. We
assume that the wind propagates into a low density medium.

After the wind material leaves the disk, it is subjected to gravity
from the black hole and radiation from the central engine. 
\citet{Proga99, Proga1} discuss one possible approach to line driven winds 
in the context of numerical modeling. To capture the most important
qualitative features of the flow and to simplify the analysis we
assume a continuum driven wind and use only Thomson opacities. This
assumption somewhat underestimates the driving force since it excludes
the contribution of line-driving. However, on the other hand, it is an
overestimate since typically a significant fraction of the central
source spectrum lies in the Compton scattering regime. To compensate
for that we introduce a factor $f_T < 1$, intended to decrease the
total luminosity $L$. The choice of a specific value of $f_T$ depends
on a particular situation being considered.

In our model, a fluid parcel of density $\rho$ leaves the disk with
the velocity of equation (\ref{vz0}), determined by the coordinate of
its launch point, and finds itself in a centrally symmetric,
conservative potential field. For Thomson scattering the radiative
force per unit volume is
\beq
\vec{F}^{rad}(r)=\frac{1}{c}\frac{\vec{r}}{|\vec{r}|}\displaystyle\int
\alpha_{\nu}F_{\nu}d\nu = \frac{n\sigma_{T,e}}{c}F\frac{\vec{r}}{|\vec{r}|},
\eeq
and the effective potential field is
\beq
U(r) = U_{rad} + U_{grav} = \frac{\rho GM_{BH}}{r}\Bigg(
\frac{f_{T}L}{L_{Edd}}-1\Bigg)=\rho\frac{\alpha}{r},
\label{u}
\eeq
where $L_{Edd}= 1.25\cdot 10^{38} \ erg/s (M_{BH}/\msol)$ is the
Eddington luminosity. Note, that here we give an expression for the
potential energy per unit volume.

\subsection{Description of the flow field}

Finding a trajectory and velocity of a fluid parcel in the outflow
reduces to the problem of the motion of a particle in a centrally
symmetric repulsive potential, described by (\ref{u}).  The trajectory
is a hyperbola \citep{Landafshitz1}, and in polar coordinates is
\beq
r(\phi)=\frac{p}{e\cos \phi -1}.
\label{traj}
\eeq
Here $\phi \in [0,\pi/2]$. Eccentricity of the orbit $e$ and parameter
of the orbit $p$ are defined as
\beq
e=1+\frac{p}{r_0}, \ \ p=2K\bigg(\frac{r_0}{r^*}\bigg)^{2n+1}r_0,
\label{ep}
\eeq
where $r_0=\lambda_0$ is the coordinate of the fluid element launch
point. The dimensionless quantity
\beq
K=\lambda^*(v^*)^2/2\alpha,
\label{K}
\eeq
is the ratio of the initial kinetic energy of a fluid particle
launched at $\lambda^*$ to its potential energy and measures the
relative importance of the disk wind vs. central radiation source.
($K$ and $n$ are the key parameters in our analysis). Equation
(\ref{traj}) can be considered as an implicit solution for $\lambda_0$
as a function of $z$ and $\lambda$. For any point $(\lambda,z)$ we can
uniquely determine the launch point of the streamline,
$\lambda_0$. Fig.~\ref{streamlines} shows a sample streamline pattern.

The velocity distribution along a streamline is
\beq
\left\{
\begin{array}{lll}
v_{\lambda}(\lambda_0,\lambda) & = & \displaystyle
\sqrt{\frac{\alpha}{\lambda_0}(e+1)}\frac{\sinh\xi}{e\cosh\xi+1}, \\
v_{z}(\lambda_0,\lambda) & = & \displaystyle
\sqrt{\frac{\alpha}{\lambda_0}(e^2-1)(e+1)}\frac{\cosh\xi}{e\cosh\xi+1}.
\end{array} \right.
\label{vstr}
\eeq
Parameter $\xi \in [0,+\infty)$ is defined by  
$\cosh \xi = (\lambda/\lambda_0) (e+1)-e$.

\subsection{Stabilty analysis}

We base our linear stability analysis for the inviscid case on the
\emph{Fj\o rtoft theorem}, which is derived from the inviscid 
limit of the Orr-Sommerfeld equation, i.e. Rayleigh equation
\citep{Yih, Maslowe}. We have studied the outflows with the values of
$-2.5 \le n \le 0.0$ and $0.05 \le K \le 2.5$. In all the regimes
studied, the disk wind is linearly stable to infinitesimal velocity
perturbations.

However a study of equation (\ref{traj}) shows that a non-linear
instability can be present in a subregion of the flow. Depending on
$n$, a streamline inclination angle at any given point $(r,\lambda)$
is either a monotonically increasing or monotonically decreasing
function of $\lambda_0$. The condition for the change in the character
of monotonicity can be obtained: using equation (\ref{traj}) we can
determine the inclination angle of a given streamline $\tan
\Theta(\lambda_0,\lambda)=dz/d\lambda.$ Using 
\beq
\frac{d}{d\lambda_0}\Big(\lim_{\lambda \rightarrow \infty}
\tan \Theta(\lambda_0,\lambda)\Big)=0,
\eeq
we find that the monotonicity changes for $n=-1/2$, i.e. the Keplerian
disk is the borderline case. This means that for $n>-1/2$, each
successive streamline is less inclined towards the abscissae axis than
the preceding one and eventually intersects all of the preceding
streamlines. For $n \leq -1/2$ the inclination angle is a
monotonically non-increasing function of $\lambda_0$ and then the
streamlines never intersect.

The intersection of two streamlines can be approximated as an
intersection of two jets of vanishing thickness whose interaction
leads to a resultant turbulent jet with an inclination angle lying
between that of the two original jets (e.g., see
\citep{Landafshitz2}). Therefore, streamline intersection causes the
biconical zone of compressible turbulence in the overall disk-wind
outflow. The extent of such a zone can be estimated: since the
streamline inclination angle is a monotonically increasing function of
$\lambda_0$, the streamline originating at the innermost radius of the
disk is the most inclined one. Therefore, it delineates the lower
boundary of the turbulent biconical zone. Its inclination angle is
\beq
\tan\Theta_{min} = \lim_{\lambda \rightarrow \infty}\frac{dz}{d\lambda}
(\lambda^*,\lambda)=2\sqrt{K+K^2}.
\label{thetamin}
\eeq
An exact shape of the upper boundary of the bicone can be found by
solving $dz/d\lambda_0 = 0$ for $\lambda_0$ as a function of $\lambda$
and then substituting it into the equation (\ref{traj}) to find the
exact dependence $z(\lambda)$ for the upper boundary. Instead, we give
an approximate expression for the inclination angle of the outer
boundary, found by analogy to equation (\ref{thetamin}):
\beq
\begin{array}{lll}
\displaystyle \tan\Theta_{max} & = & \lim_{\lambda \rightarrow \infty}
\frac{dz}{d\lambda}(\lambda_0,\lambda) \\
& = & 2\sqrt{K\Big(\frac{\lambda_0}{\lambda^*}\Big)^{2n+1}
+K^2\Big(\frac{\lambda_0}{\lambda^*}\Big)^{2(2n+1)}}.
\label{thetamax}
\end{array}
\eeq
This defines the basic geometry of the bicone: the \emph{inclination
angle} $\Theta = (\Theta_{max}+\Theta_{min})/2$ and the
\emph{divergence angle} $\Delta \Theta = \Theta_{max}-\Theta_{min}$.
See Figure~\ref{streamlines}.

Typically, the bicone has a divergence angle $\Delta\Theta$ that
slowly increases with radius. Therefore equation (\ref{thetamax}),
which gives the limiting inclination angle of the streamline
originating from the point $\lambda_0$, depends on the choice of such
a particular streamline. However, the increase in the divergence angle
is typically small. Moreover, a particular streamline used to
determine $\Theta_{max}$ can be determined independently from the
properties of the accretion disk or the estimated extent of the
outflow.

Three other effects will enter in a more detailed hydrodynamic
treatment that influence the divergence angle of the bicone. An effect
which tends to make the divergence angle smaller than that determined
using equations (\ref{thetamin}) and (\ref{thetamax}) arises because
the resultant of two intersecting jets lies at an intermediate angle
between the original two. Each successive interaction of resultant
jets would then tend to narrow the overall conical structure. In
addition, the change in the optical depth of the outflow after the
formation of the turbulent bicone would shade the outer parts of the
outflow. This will make the bicone narrower by increasing the
inclination angle of the outer streamlines due to the reduced
radiation pressure. In competition with both of these above effects
is the turbulence itself, which could broaden the bicone.

The initial density fluctuations in the turbulent biconical zone
cannot themselves represent the BELR clouds; the density and
temperature contrasts are too small compared to the ambient warm
highly ionized gas (WHIM in the nomenclature of \citet{Elvis}).
However, thermal instability in such a medium can transform initial
turbulent overdensities into the BELR clouds (see \citep{Burkert,
Hennebelle99} for the discussion of two-phase medium formation via
linear thermal instability in optically thin regions). The formation
of the compressible turbulent bicone from intersecting flows can
produce inhomogeneities with density contrast up to 4 in the adiabatic
case and higher if one allows for the possibility of radiative
cooling. Such high density contrasts (compared to the linear density
fluctuations discussed by \citet{Burkert}) can serve as a seed for
thermal instability. The instabilty proceeds in the nonlinear regime
from the onset. Rapid cooling would then produce clumps. The maximum
clump density reached at a given scale is determined by how much
cooling can occur during an eddy turnover time on that scale, i.e.
during an average clump survival time.

This mechanism can lead to a dynamical equilibrium described by the
balance between in situ formation, destruction, and re-formation of
clumps comoving with the ambient flow. As long as any given clump
survives long enough to reprocess radiation, problems of clump
confinement against thermal conduction and destruction via high
velocity differences between the clumps and the ambient medium are
eliminated. The dynamical equilibrium of the turbulent flow, ensures a
constant number density of BELR clouds in the turbulent biconical zone
over observation durations.

Finally, there is a site where a linear instability of the
Kelvin-Helmholtz type might operate, namely the shear layer between
the disk wind outflow and the infalling (or stationary) circumpolar
material, that fills the cone of the outflow. Simulations of
\citet{Proga1} show the formation of clumpy structure in 
that boundary shear layer. They attribute it to the Kelvin-Helmholtz
instability. However, in regimes for which the outflow develops the
turbulent bicone, such linear instability will be completely
overwhelmed by the nonlinear instability induced by the intersection
of streamlines. In cases when the bicone is not formed, the outflow
density decreases from the maximum values near the plane of the disk
to nearly the values of the circumpolar infalling material in the
upper parts of the outflow with the density transition occurring over
the large range of angles. BLR clouds are not expected in this case.
\footnote{Note, that even in the cases when the turbulent bicone is 
absent the upper boundary of the outflow is still roughly defined by
the streamline launched at $\lambda^*$.}.

\section{DISCUSSION}

We considered the linear and nonlinear stability of a disk outflow and
found that it is linearly stable to infinitesimal velocity
perturbations. However, when $n>-1/2$ and a supereddington radiation
source is present, nonlinear instability leads to the formation of a
biconical zone of compressible turbulence. High density contrasts in
this zone may trigger thermal instability and lead to the further
condensation of clumps into BELR clouds. Turbulence plays a key role
in establishing the dynamical equilibrium which maintains a steady
cloud number density, even though each cloud is short lived.

\begin{figure}
\plotone{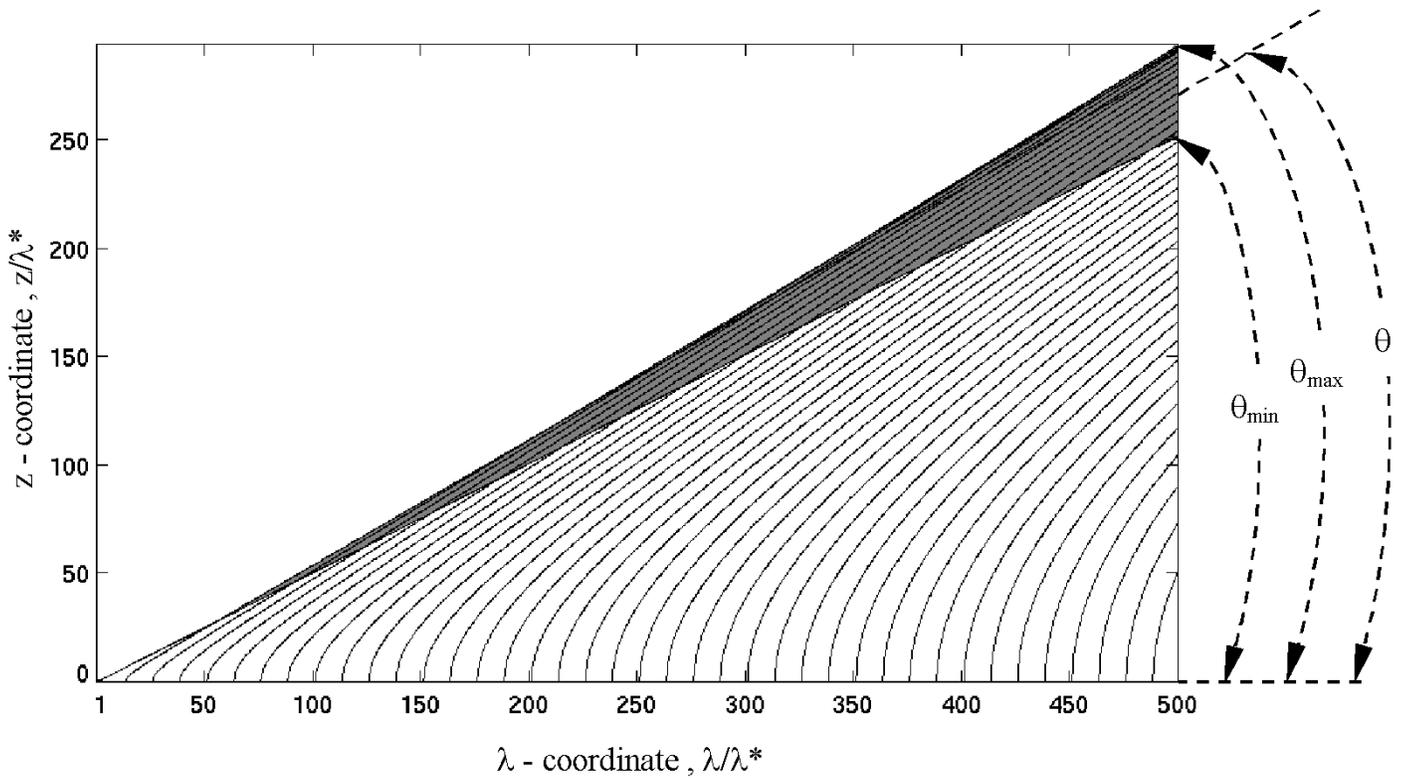}
\caption{Sample streamline pattern for a disk wind outflow. Shown is the case 
$n=-0.45$ and $K=0.06$. Shaded region is the turbulent bicone formed
by the intersection of streamlines. Values of the angles shown are
discussed in the text.
\label{streamlines}}
\end{figure}

In the regimes when the outflow can develop the turbulent bicone, the
geometry is remarkably similar to that described by \citet{Elvis}.
For a disk outflow with $K \approx 0.05$ and an initial launch
velocity profile slightly flatter than Keplerian, namely $n=-0.45$,
the turbulent bicone inclination angle is $\Theta \approx 28^o$ with
the divergence angle $\Delta\Theta \approx 7^o$. This is in
quantitative and qualitative agreement with \citet{Elvis}: (1) no
absorbers are along the line of sight passing above the outflow; (2)
along the line of sight passing through the turbulent bicone broad
absorption and emission line features will be observed; (3) narrow
absorption lines (NAL) will be seen along sight lines that fall inside
of the bicone. Figure~\ref{streamlines} shows the geomtry of the
outflow and the emerging bicone with the angle values found here.

The outflow geometry in our picture depends only on dimensionless
quantities $K$ and $n$. However, in addition to $\Theta$ and
$\Delta\Theta$ determined above, we can also infer the inner scale by
matching observed cloud velocities. We assume the same value of $n$
and $K$ as in the previous paragraph. We take $M_{BH} =10^9\ \msol$
and broad absorption line velocity $v_{BAL}=10^4$ \kms (the velocity
observed when the line of sight passes directly along the bicone
towards the central source). In our model $v_{BAL}$ is the terminal
velocity at infinity of material launched with the largest initial
velocity and potential, i.e. at the point $\lambda_0=\lambda^*$.
Setting $\lambda_0=\lambda^*$ in (\ref{vstr}) and using the
expressions for $v_{\lambda}(\lambda^*,\lambda)$ and
$v_{z}(\lambda^*,\lambda)$ in the limit $\lambda \rightarrow \infty$,
we find 
\beq
\lambda^* =\frac{2\alpha(1+K)}{v_{BAL}^2}.
\label{lstar}
\eeq
Assuming $L=1.5L_{Edd}$ and $f_{T}=0.7$ the corresponding value of
$\lambda^* =\lambda_0 \approx 1000 \ a.u.$, which matches
\citet{Elvis}. The corresponding maximum launch velocity is $\approx
2000$ \kms. Therefore, the observed velocity in the NAL region is
$\sim 1000$ \kms.

In principle, the paradigm presented here applies not only to luminous
AGN but to any source with a centrally symmetric potential, a
sufficiently luminous central radiation source, and a disk wind,
e.g. disk winds in young stellar objects.

We have not discussed the role of magnetic fields, disk rotation,
metallicity, wind density fall off, or the physics of line driving.
These should be considered in future work.

\acknowledgements

The authors thank V. Pariev for extremely important comments and
suggestions. AYP thanks N. Murray and M. Elvis for valuable
discussions. AYP and AF acknowledge support from NSF grant AST-9702484
and NASA grant NAG5-8428. EGB acknowledges support from DOE grant
DE-FG02-00ER5460. The authors acknowledge DOE support from the
Laboratory for Laser Energetics.

\clearpage

\begin{thebibliography}{}
%
\bibitem[Arav \ea (1994)]{Arav94}
Arav, N., Li, Z.-Y., Begelman, M. C. 1994, \apj, 432, 62
%
\bibitem[Arribas \ea (1996)]{Arribas96}
Arribas, S., Mediavilla, E., Garcia-Lorenzo, B. 1996, \apj, 463, 509
%
\bibitem[Bottorff \& Ferland (2000)]{Bottorff1}
Bottorff, M. C., Ferland, G. J. 2000, \mnras, 316, 103
%
\bibitem[Bottorff \& Ferland (2001)]{Bottorff2}
Bottorff, M. C., Ferland, G. 2001, \apj, 549, 118
%
\bibitem[Bottorff \ea (2000)]{Bottorff3}
Bottorff, M., Ferland, G., Baldwin, J., Korista, K. 2000, \apj, 542, 644
%
\bibitem[Burkert \& Lin (2000)]{Burkert}
Burkert, A., Lin., D. N. C. 2000, \apj, 537, 270
%
\bibitem[Celotti \ea (1992)]{Celotti} 
Celotti, A., Fabian, A. C., Rees, M. J. 1992, \mnras, 255, 419
%
\bibitem[Elvis (2000)]{Elvis}
Elvis, M. 2000, \apj, 545, 63
%
\bibitem[Hennebelle \& P\'erault (1999)]{Hennebelle99}
Hennebelle, P., P\'erault, M. 1999, \aap, 351, 309
%
\bibitem[K\"onigl \& Kartje (1994)]{Konigl94}
K\"onigl, A., Kartje, J. F. 1994, \apj, 434, 446
%
\bibitem[Kuncic \ea (1996)]{Kuncic1}
Kuncic, Z., Blackman, E. G., Rees, M. J. 1996, \mnras, 283, 1322
%
\bibitem[Kuncic \ea (1997)]{Kuncic2}
Kuncic, Z., Celotti, A., Rees, M. J. 1997, \mnras, 284, 717
%
\bibitem[Landau \& Lifshitz (1959)]{Landafshitz2}
Landau, L. D., Lifshitz, E. M. 1959, Fluid Mechanics
(Reading : Addison-Wesley)
%
\bibitem[Landau \& Lifshitz (1976)]{Landafshitz1}
Landau, L. D., Lifshitz, E. M. 1976, Mechanics (Oxford, New York : Pergamon Press)
%
\bibitem[Maslowe (1985)]{Maslowe}
Maslowe, S. A. 1985, in Hydrodynamic Instabilities and Transition to Turbulence,
ed. H.L. Swinney, J.P. Gollub (2nd ed.;Springer-Verlag), 181
%
\bibitem[Mathews \& Ferland (1987)]{Mathews} 
Mathews, W. G., Ferland, G. J. 1987, \apj, 323, 456
%
\bibitem[Murray \& Chiang (1995)]{Murray1}
Murray, N., Chiang, J. 1995, \apjl, 454, L105
%
\bibitem[Murray \ea (1995)]{Murray2}
Murray, N., Chiang, J., Grossman, S. A., Voit, G. M. 1995, \apj, 451, 498
%
\bibitem[Pelletier \& Pudritz (1992)]{Pudritz92}
Pelletier, G., Pudritz, R. E. 1992, \apj, 394, 117
%
\bibitem[Pier \& Voit (1995)]{Pier95}
Pier, E. A., Voit, G. M. 1995, \apj, 450, 628
%
\bibitem[Pietrini \& Torricelli-Ciamponi (2000)]{Pietrini}
Pietrini, P., Torricelli-Ciamponi, G. 2000, \apj, 363, 455
%
\bibitem[Poludnenko \ea (2002)]{Poludnenko}
Poludnenko, A. Y., Frank, A., Blackman, E. G. 2002, \apj, in press
%
\bibitem[Proga \ea (1999)]{Proga99}
Proga, D., Stone, J. M., Drew, J. E. 1999, \mnras, 310, 476
%
\bibitem[Proga \ea (2000)]{Proga1}
Proga, D., Stone, J. M., Kallman, T. R. 2000, \apj, 543, 686
%
\bibitem[Ruiz \ea (2001)]{Ruiz01}
Ruiz, J. R., Crenshaw, D. M., Kraemer, S. B., Bower, G. A., Gull, T. R., 
Hutchings, J. B., Kaiser, M. E., Weistrop, D. preprint (astro-ph/0108521)
%
\bibitem[Rees (1987)]{Rees87}
Rees, M. J. 1987, \mnras, 228, 47p
%
\bibitem[Urry \& Padovani (1995)]{Urry95}
Urry, C. M., Padovani, P. 1995, \pasp, 107, 803
%
\bibitem[Yih (1969)]{Yih}
Yih, C. S. 1969, Fluid Mechanics (2d ed.; McGraw-Hill, New York 1969)

\end{thebibliography}
\end{document}